# Snowball: Strain aware gene assembly of Metagenomes

I. Gregor[1,2], A. Schönhuth[3,*], A. C. McHardy[1,2,*]

[1]Department of Algorithmic Bioinformatics, Heinrich-Heine-University Düsseldorf, Düsseldorf, Germany
[2]Computational Biology of Infection Research, Helmholtz Center for Infection Research, Braunschweig, Germany
[3]Centrum Wiskunde & Informatica, Amsterdam, The Netherlands
[*]Joint last authors

## Abstract

Gene assembly is an important step in functional analysis of shotgun metagenomic data. Nonetheless, strain aware assembly remains a challenging task, as current assembly tools often fail to distinguish among strain variants or require closely related reference genomes of the studied species to be available. We have developed *Snowball*, a novel strain aware and reference-free gene assembler for shotgun metagenomic data. It uses profile hidden Markov models (HMMs) of gene domains of interest to guide the assembly. Our assembler performs gene assembly of individual gene domains based on read overlaps and error correction using read quality scores at the same time, which result in very low per-base error rates. The software runs on a user-defined number of processor cores in parallel, runs on a standard laptop and is freely available for installation under Linux or OS X on:
https://github.com/algbioi/snowball/wiki

## 1. Introduction

Metagenomics is the functional or sequence-based analysis of microbial DNA isolated directly from a microbial community of interest (Kunin *et al.*, 2008; Riesenfeld *et al.*, 2004). This enables analysis of microorganisms that cannot be cultivated in a laboratory. After the DNA is isolated, it is sequenced using a high-throughput sequencing platform, which results in a large dataset containing short sequences called reads, where it is unknown for a read from which genome sequence it oritinates. Given such sequenced shotgun metagenomic data, i.e. a dataset of short reads that originate from several genome sequences of distinct strains, the goal of the gene assembly is to reconstruct coding sequences of the individual strains contained in the dataset (Fig. 1).

For many purposes, including functional analysis of the metagenomic data, it is sufficient and therefore convenient to assemble only the coding sequences of the strains. It has also been shown that genes assemble well (Kingsford *et al.*, 2010) even when only short reads are available. Moreover, metagenomic data consist mainly of prokaryotic species. As usually more than 85% of prokaryotic genomes are coding sequences (Cole and Saint-Girons, 1999); gene assembly enables to recover large parts of the respective genomes. Gene assembly is therefore an important step in the analysis of shotgun metagenomic data.

Importantly, strain awareness is an essential goal in assembling metagenomes, since it enables us to study gene variation among strains of metagenomic species. Nonetheless, the assembly of closely related strains remains a challenging task, since current assemblers often fail to distinguish among individual strain variants and similar sequences of distinct strains get assembled into consensus sequences, which removes the strain variation information.

Tools that enable strain variant reconstruction often rely on the availability of closely related reference genomes of the studied species (Zagordi *et al.*, 2011; Töpfer *et al.*, 2014; Ahn *et al.*, 2015), where reads are first mapped onto a reference genome, using a read mapping tool, e.g. *BWA* (Li and Durbin, 2009), strain variants are then identified through a reference guided strain aware assembly. As metagenome samples originating from novel environments typically consist of novel species without reference genomes available, there is a need for new reference-free approaches.

Tools that are often used for de novo metagenome assemblies are *IDBA-UD* (Peng *et al.*, 2012), *MetaVelvet* (Namiki *et al.*, 2012), *Velvet* (Zerbino and Birney, 2008) or *SOAPdenovo2* (Luo *et al.*, 2012). All these tools are $k$-mer based, i.e. they transform reads into overlapping $k$-mers from which De Bruijn graphs are built, where paths in the graph correspond to the assembled contigs. This general approach, however, often fail to distinguish among strain variants. There has been recent debate on $k$-mer based approaches using De Bruijn graphs in strain aware assembly. In particular, when low frequencies strains are involved, since the frequencies of the low abundant strains are on the order of magnitude of the sequencing error rates. This leads to unpleasant interference in the error-correction step of the $k$-mer based approaches, as low frequencies strains are often removed along with sequencing errors. There has also been recent evidence that shorter genomes can be assembled through overlap graph based approaches using short reads (Simpson and Durbin, 2012). It was also shown that one can perform strain aware assembly through iterative construction of overlaps graphs (Töpfer *et al.*, 2014). For the gene assembly of metagenomic data, the *SAT* assembler (Zhang *et al.*, 2014) can be employed. First, it assigns reads to gene domains of interest based on profile hidden Markov models (HMMs) (Eddy, 2011; Finn *et al.*, 2014) of the respective gene domains. Then, for each gene domain, separately, it builds overlap graphs based on the read overlaps, where the paths in the graphs correspond to the assembled contigs. As the *SAT* assembler does not implement any sophisticated error-correction strategy, it also often fails to distinguish among gene strain variants (Section 3.1). For the reconstruction of only the 16S genes, which is often used for phylotyping, *REAGO* (Yuan *et al.*, 2015) can be employed, however its use for the functional analysis of metagenomic data is limited.

The current sequencing technologies still produce relatively short erroneous reads, thus it is very difficult to distinguish sequencing errors from genuine strain variation (Laehnemann *et al.*, 2015). Therefore, reference-free strain reconstruction of the full-length sequences of individual strains is considered to be currently infeasible, since there is not enough information contained in the sequenced data. This also suggests that the library preparation is still to be improved.

Here, we present *Snowball*, a novel method for strain aware and reference-free gene assembly of metagenomes. It uses profile HMMs of gene domains of interest as an input to guide the assembly. The HMM profile-based homology search is known to be capable of finding remote homology, including large number of substitutions, insertions and deletions, whereas simple read mapping onto a reference genome can find only very closely related homologs (Zhang *et al.*, 2014). Our method is reference-free, thereby allowing for strain aware gene assembly of novel species. We have developed a novel algorithm that performs gene assembly based on read overlaps, which allows to correct errors by making use of the error profiles that underlie the overlapping reads. The consequences are twofold. First, we obtain contigs affected by only very low per-base error rates. Second, since, this way, we determine which reads stem from identical segments based on a statistically sound model, we can safely distinguish between sequencing errors and strain-specific variants, even of very low abundant strains. We consider these two features the main improvements over the currently available assemblers. A new algorithm facilitates an appropriate implementation of the underlying methodical concepts. To our knowledge, *Snowball* is the first tool that enables us to distinguish among individual gene strain variants in metagenomes for a large set of gene domains without using reference genomes of related species.

We have assessed the performance of *Snowball* using ten simulated datasets, each containing three closely related *E.Coli* strains. The results confirm that the strength of our tool is its very low per-base error, due to the incorporated error-correction. Moreover, it produced substantially longer contigs and recovered larger part of the simulated reference data than the *SAT* assembler (Section 3). *Snowball* is implemented in Python, runs on a user-defined number of processor cores in parallel, runs on a standard laptop, is freely available and can be installed under Linux or OS X.

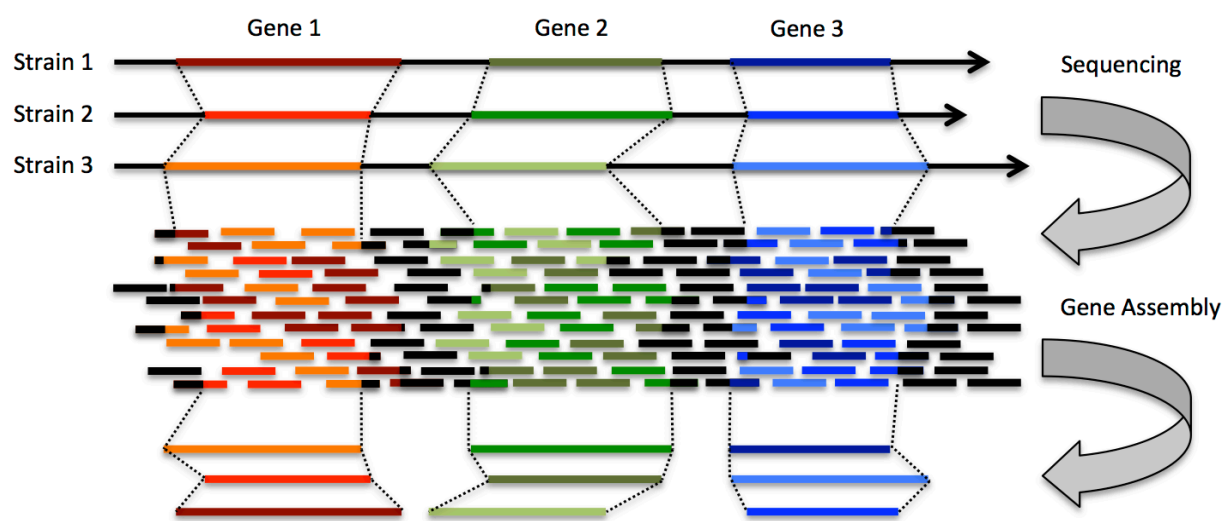

**Figure 1.** An example of the gene assembly problem.
In this example, the metagenomic community consists only of three distinct strains. Non-coding regions of the strain sequences are black, whereas coding regions are red, green and blue for genes 1, 2 and 3. Genes 1–3 are present in all three strains, although the location and gene sequences differ for distinct strains. The sequencing step results in a collection of short reads. Note that after the sequencing step, the origin of reads denoted by colors in the figure is not known in the subsequent gene assembly step. Given a dataset containing all the short reads, the ultimate goal of the gene assembly is to determine the individual strain specific sequences of the genes.

## 2. Methods

The input of *Snowball* are two FASTQ files containing Illumina self-overlapping paired-end reads of closely related strains, the corresponding insert size used for the library preparation and profile HMMs of gene domains of interest. We have thoroughly tested *Snowball* using simulated Illumina HiSeq 2500 paired-end reads generated by the *ART* read simulator (Huang *et al.*, 2012) with 150bp read length and 225bp mean insert size. In this setting, the average length of the self-overlaps of the read ends is 75bp and the length of a consensus read that originates by joining of the self-overlapping read ends is 225bp on average (Fig 2, Section 3.4). The output is a FASTA or a FASTQ file containing annotated assembled contigs. For each contig, the annotation contains the name of a respective gene domain to which a contig belongs, coordinates of the coding sub-sequence within a contig sequence, coverage and quality score for each contig position. The coverage and quality score information can be used for subsequent quality filtering yielding less or shorter contigs of higher quality.

Our method consists of the following steps:
- [Consensus read reconstruction]
  Self-overlapping paired-end reads are joined into longer consensus reads (Section 2.1).
- [Assignment of consensus reads to gene domains]
  Profile HMMs of selected gene domains are employed to assign consensus reads to the respective gene domains, where one consensus read is assigned to at most one gene domain (Section 2.2).
- [Assembly of consensus reads into contigs]
  For each gene domain, in parallel, consensus reads are assembled into contigs (Sections 2.3–2.5). In the assembly step, consensus reads are iteratively joined into longer and error-corrected super-reads based on the consensus read overlaps. The super-reads are then output as annotated contigs, where a super-read represents a sequence that originates by joining of at least two consensus reads into a longer sequence.

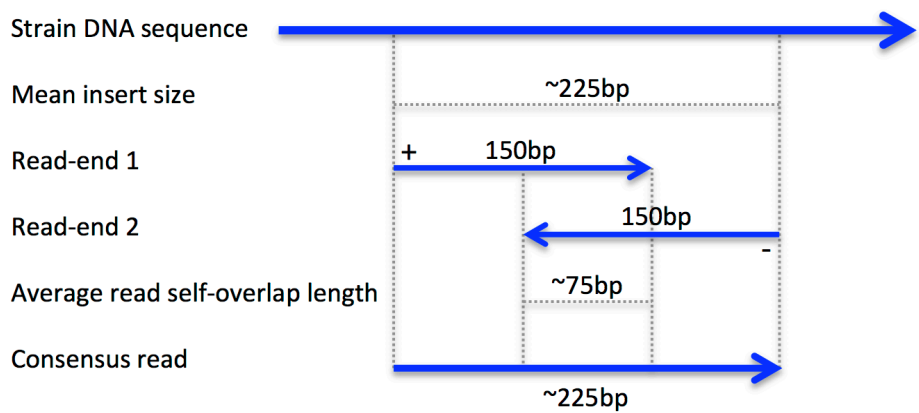

**Figure 2.** An example of a self-overlapping paired-end read.
Illumina HiSeq 2500 paired-end read consists of two 150bp read ends, one on the positive strand (+) and one on the negative strand (-). In our example, the mean insert size (225bp) is smaller than two times the read end length (2 * 150bp), therefore the paired-end reads are self-overlapping with 75bp overlap length on average. Such a self-overlapping read can be joined into a consensus read of 225bp length on average.

## 2.1. Joining self-overlapping paired-end reads

Self-overlapping paired-end reads are joined into longer error-corrected consensus sequences. The use of a library containing self-overlapping paired-end reads is a powerful strategy for an initial error-correction (Schirmer *et al.*, 2015), which has been employed in e.g. *ALLPATHS* (Butler *et al.*, 2008). Given the mean insert size, we determine the self-overlap that results in the minimum hamming distance between the overlapping ends of a paired-end read. A base with a higher quality score is chosen at a position within the overlap that contains mismatching bases for the respective position of the resulting consensus read sequence (Fig. 3). As the substitution error rate of the Illumina reads increases towards the ends of the paired-end reads (Laehnemann *et al.*, 2015), this step results in longer consensus reads with overall lower substitution error, where the overlapping regions are almost error-free. It is also an efficient read quality filtering step, as the paired-end reads that cannot be joined, due to high substitution error rate, an insertion or a deletions within the overlapping region, are filtered out. For instance, by joining of the 150bp paired-end Illumina HiSeq 2500 self-overlapping reads with 225bp mean insert size results in consensus reads of length 225bp on average. While the default error profile of the *ART* read simulator (Huang *et al.*, 2012) yields 150bp paired-end reads with ~2.38% substitution error, the joined consensus reads have only ~1.08% substitution error. These longer, error-corrected consensus reads with low substitution error rate are convenient building blocks to start with in the subsequent steps of our method.

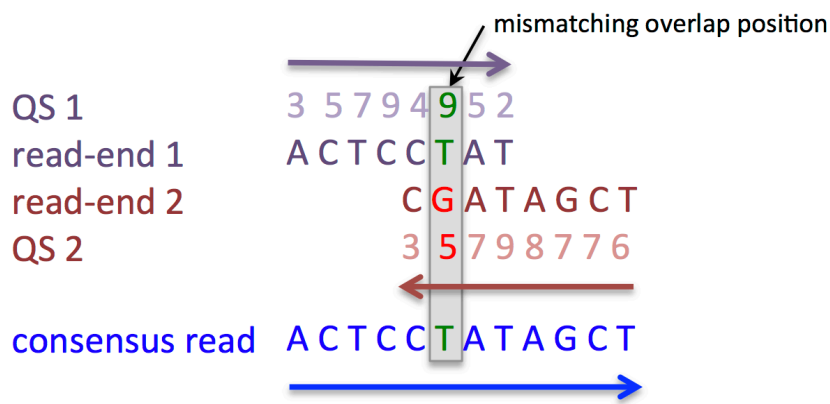

**Figure 3.** Joining of self-overlapping reads example.
The figure depicts a simplified example of a consensus read reconstruction. At the mismatching overlap position, read-end 1 has *T* with quality score 9, while read-end 2 has *G* with quality score 5. The resulting consensus read will have *T* at the respective position, since *T* is supported by a higher quality score than *G*.

## 2.2. Assigning reads to gene domains

Consensus reads are annotated using profile HMMs of gene domains of interest and assigned to respective gene domains (Fig. 4). By default, we use the Pfam-A (Finn *et al.*, 2014) (version 27) profile HMMs of 14,831 gene domains and *AMPHORA 2* (Wu and Scott, 2012) profile HMMs of 31 bacterial ubiquitous single-copy genes that are often used for phylotyping. A profile HMM of a gene domain is a probabilistic model representing a multiple sequence alignment of representative gene sequences belonging to a particular gene domain. The model can be used to annotate a query sequence (e.g. a consensus read). The annotation mainly consists of a score, start/stop positions within a query sequence and HMM start/stop coordinates. The score roughly corresponds to a probability that a query sequence belongs to the particular gene domain, i.e. if the score is high for a query sequence then it is very probable that it belongs to the respective gene domain. The start/stop positions within a query sequence define a sub-sequence of a query sequence that was identified to belong to the gene domain. The HMM start/stop coordinates correspond to the estimated coordinates of the query sub-sequence within the multiple sequence alignment of the respective profile HMM.

Each consensus read is translated into 6 protein sequences using all 6 reading frames (i.e. also considering the reverse complementary sequences). The *hmmsearch* command of the *HMMER 3* (Eddy, 2011) software is used to annotate the protein sequences. For each consensus read, only the reading frame with the highest score is considered. A consensus read is assigned to at most one gene domain to which it was queried with the highest score. Consensus reads with low scores (i.e. lower than default value: 40) are filtered out and not considered in the subsequent steps. If a protein sequence corresponding to a reverse complementary consensus read sequence was annotated, the corresponding reverse complementary DNA sequence of a respective consensus read is considered in the next steps. The coding DNA sub-sequence of a consensus read sequence is denoted as a coding region. The start and end HMM coordinates within a respective profile HMM are stored as part of the consensus read annotation.

As a result of this step, consensus reads are annotated and assigned to "bins" representing individual gene domains, where one consensus read is assigned to at most one gene domain. Gene domains are building blocks of individual genes. Therefore, a "bin" does not only contain consensus reads belonging to gene variants of individual strains. It can also contain different genes of one strain, several copies of one gene of one strain or even "broken" gene copies.

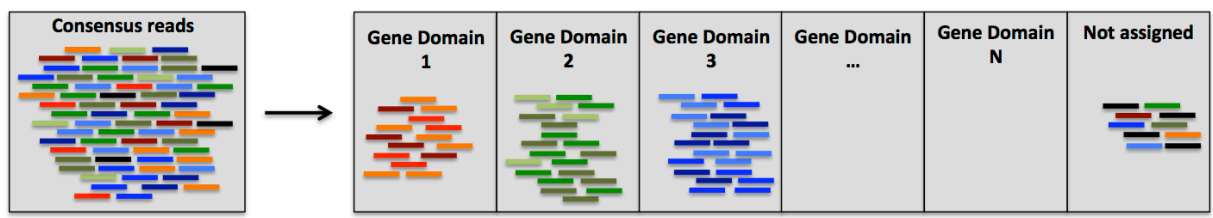

**Figure 4.** Assignment of consensus reads to gene domains.
Consensus reads are assigned to individual gene domains using profile HMMs, where consensus reads that cannot be assigned to any of the gene domains with sufficient confidence remain unassigned. A consensus read is assigned to at most one gene domain.

## 2.3. Consensus sequence representation

We represent consensus sequences, i.e. consensus reads and super-reads using probability matrices. A super-read is a longer error-corrected sequence that originates by joining of overlapping consensus reads (or consensus reads with super-reads) in the "*Snowball*" algorithm (Section 2.5). Illumina paired-end reads are stored in FASTQ files along with the corresponding quality scores (Fig. 5a). A quality score for a read position represents a probability that a base was sequenced correctly, i.e. it represents a probability that a particular base is at a respective position in the FASTQ file (Fig. 5b). The complement probability represents a probability that a different base is at a respective position. A probability that different base *X* is at a particular position corresponds to one third of the complement probability. Note that these probabilities are only estimates provided by the Illumina sequencing platform.

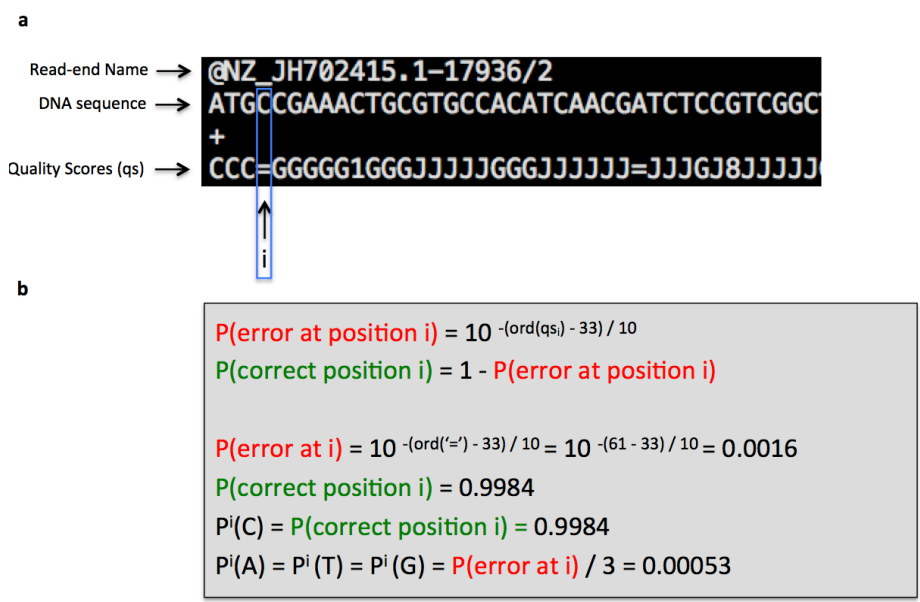

**Figure 5.** FASTQ file data representation.
(Panel a) depicts example of a read end representation in a FASTQ file. The entry consists of a read end name, DNA sequence of the respective end of a paired-end read and a quality score for each position of the DNA sequence. (Panel b) explains the meaning of the quality scores. From quality score $qs_i$ at position *i*, we compute the probability that position *i* was correctly sequenced, where the *ord* function assigns an ASCII number to an input ASCII character. The probability that base *C* is at position *i* is equal to the probability that position *i* was sequenced correctly. Probability of *A*, *T*, or *G* being at position *i* is equal to the probability that position *i* was sequenced incorrectly divided by three.

A probability matrix represents a consensus sequence, where each sequence position is represented by a probability distribution over DNA bases {A, C, T, G}. An example of a probability matrix corresponding to a consensus sequence of two overlapping sequences is depicted in (Fig 6). At a particular position within a consensus sequence, we compute the expected probability of a base as the average probability of the respective base probabilities of the individual reads covering the position. The individual base probabilities are derived from the quality scores (Fig. 5). Let $R$ be the set of all read ends that were joined into consensus sequence $c$ and cover position $p_c$ within $c$. For a read end $r \in R$, let $p_r$ be the position within $r$ that corresponds to position $p_r$ within the consensus sequence $c$. The probability of a base $X \in \{A, C, T, G\}$ being at position $p_c$ within the consensus sequence $c$ is:

$$P^{p_c}(X) = \frac{1}{|R|} \sum_{r \in R} P_r^{p_r}(X)$$

A probability matrix represents a consensus DNA sequence, where the base with the highest probability at a particular position is the base of the consensus DNA sequence at the respective position.

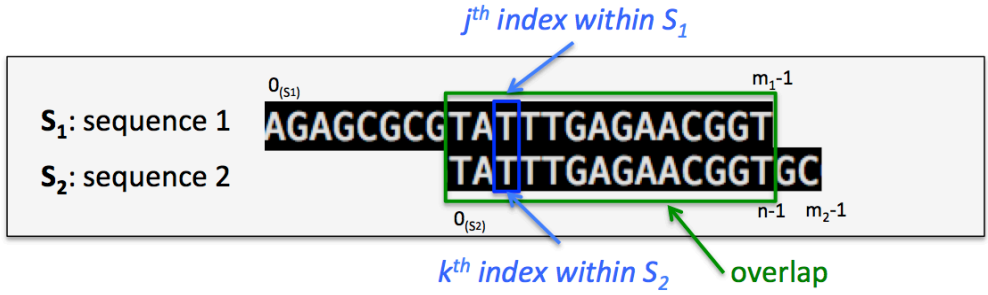

| | $S_1$ | | | $S_1 \cap S_2$ | | | $S_2$ |
|---|---|---|---|---|---|---|---|
| P(A) | $P_1^0(A)$ | ... | ... | $[P_1^j(A) + P_2^k(A)]/2$ | ... | ... | $P_2^{m_2-1}(A)$ |
| P(C) | $P_1^0(C)$ | ... | ... | $[P_1^j(C) + P_2^k(C)]/2$ | ... | ... | $P_2^{m_2-1}(C)$ |
| P(T) | $P_1^0(T)$ | ... | ... | $[P_1^j(T) + P_2^k(T)]/2$ | ... | ... | $P_2^{m_2-1}(T)$ |
| P(G) | $P_1^0(G)$ | ... | ... | $[P_1^j(G) + P_2^k(G)]/2$ | ... | ... | $P_2^{m_2-1}(G)$ |

**Figure 6.** Probability matrix example.
An example of a probability matrix construction, where two overlapping sequences are joined into a consensus sequence and represented as a probability matrix. The superscripts of individual probabilities correspond to positions within respective sequences. The subscripts correspond to either sequence $S_1$ or $S_2$. The probability arguments are DNA bases {A, C, T, G}.

### 2.4. Overlap probabilities and error correction

The computation of overlap probabilities of two overlapping sequences is an essential part of the "*Snowball*" algorithm. Given two overlapping sequences $S_1$ and $S_2$ as depicted in (Fig. 6), where $n$ is the length of the overlapping region, the overlap probability at position $i \in [0, ..., n-1]$ is computed as:

$$P_{overlap}^i = \sum_{X \in \{A,C,T,G\}} P_1^i(X) * P_2^i(X)$$

Where, $P_1^i(X)$ is a probability that sequence $S_1$ has base $X$ at overlap position $i$; probability $P_2^i(X)$ is defined analogously for sequence $S_2$. The overall overlap probability of $S_1$ and $S_2$ is the product of individual position overlap probabilities normalized by overlap length $n$ (Töpfer *et al.*, 2014):

$$P_{overlap} = \sqrt[n]{\prod_{i \in [0,..,n-1]} P_{overlap}^i}$$

As a score that represents the "expected length" of an overlap, taking into account the individual overlap position probabilities, we compute the expected number of correct positions within the overlap as:

$$lengthExpected = \sum_{i \in [0,..,n-1]} P_{overlap}^i$$

A single overlap score that enables us to rank different sequence overlaps is computed as a product of the overall overlap probability and the expected overlap length:

$$scoreOverlap = P_{overlap} * lengthExpected$$

The overlap score penalizes both overlaps with low overlap probability and short overlaps, since long overlaps with high overlap probability are required. The minimum required expected length of an overlap represents the support for the overlap probability, as the overlap probability is based only on the bases within the overlap, therefore the number of the bases outside of the overlap should remain as small as possible, since we cannot make any statement about the bases outside of the overlap.

In the "*Snowball*" algorithm, consensus reads are iteratively joined into longer super-reads based on the overlap probabilities, expected overlap lengths and the overlap scores (Section 2.5). By default, two sequences $S_1$ and $S_2$ can be joined into a consensus sequence if the overall overlap probability is at least *0.8* and the expected length of the overlap is at least *0.5 \* min(length(S1), length(S2))*. The high overall overlap probability ensures that the overlap consists of mostly matching positions, that there are no mismatching positions with high quality scores and that mismatches are allowed only at positions with low quality scores. For datasets with overall high quality scores, the minimum overlap probability parameter can be increased to 0.9 or 0.95. In the "*Snowball*" algorithm, when a consensus sequence can be joined with multiple consensus sequences with sufficient overlap probability and expected overlap length, it is joined with the sequence with which it has the highest overlap score.

## 2.5. The "*Snowball*" algorithm

For each gene domain, the "*Snowball*" algorithm iteratively joins consensus reads into longer error-corrected super-reads. The input of the algorithm are annotated consensus reads of a particular gene domain represented via probability matrices (Sections 2.1–2.3). The resulting super-reads are output as annotated contigs. Note, that the method can be highly parallelized, since the "*Snowball*" algorithm runs for each gene domain separately.

Consensus reads are first sorted in an increasing order according to the HMM start coordinates, that denote an estimated start position of a consensus read within the multiple sequence alignment of the profile HMM. This layout suggests which pairs of consensus reads are likely to have an overlap (Fig. 7), where consensus reads that are next to each other are likely to have longer overlaps than other pairs of consensus reads.

As a starting point of the algorithm, we choose a consensus read with the largest sum of overlap lengths with other consensus reads and put it into the *working set*. The reason for this choice is that such a consensus read is within the highest coverage of the alignment corresponding to the respective profile HMM, where highly covered regions are likely to be covered by reads originating from similar but distinct genomes. Therefore, the chosen consensus read is very likely to overlap with consensus reads originating from different gene variants.

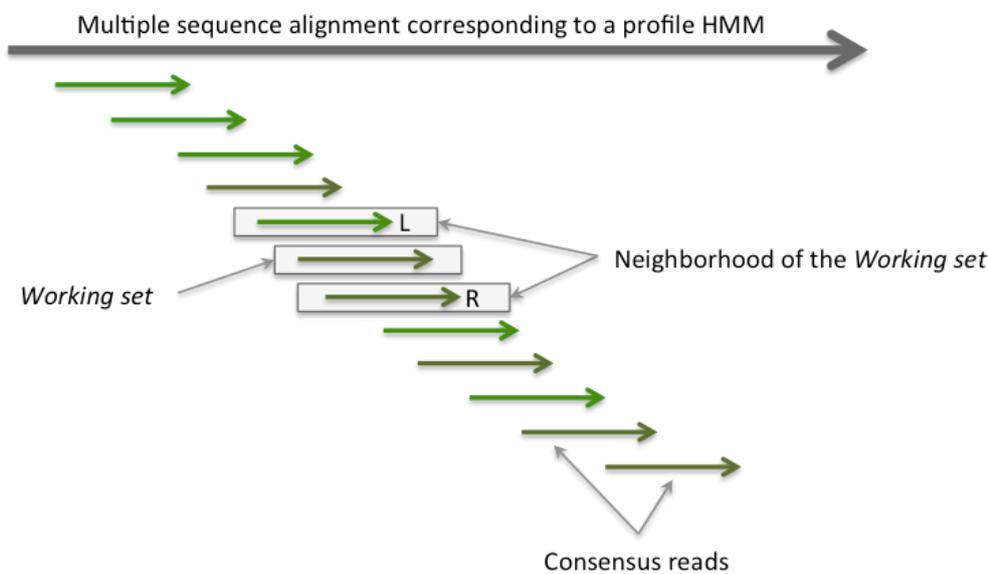

**Figure 7.** Initial layout of consensus reads.
Consensus reads sorted according to the HMM start coordinates. In the neighborhood of the consensus read, that is in the *working set*, there are two closest consensus reads, one on the left ($L$) and one on the right ($R$).

The main idea of the algorithm is that we iteratively try to extend consensus sequences from the *working set* into longer consensus sequences by joining them with consensus reads that are in their neighborhood, considering the consensus read layout (Fig. 7). In one iteration, a consensus read from the neighborhood (i.e. $L$ or $R$) is joined with one of the consensus sequences from the *working set* or two consensus reads (i.e. $L$ and $R$) that are in the neighborhood of the *working set* are added to the *working set*. A consensus read and a consensus sequence are joined only if they have a sufficient overlap as defined in (Section 2.4). If more than one overlap of a consensus read from the neighborhood (i.e. $L$ or $R$) and a consensus sequence from the *working set* is sufficient, the pair that has the highest overlap score is chosen. If there is no sufficient overlap between a consensus sequence from the *working set* and a consensus read $L$ or $R$ in the neighborhood, both consensus reads are added to the *working set* as they are likely to originate from distinct gene variants than the gene variants already represented in the *working set*. The pseudo code of the algorithm:

1. Input: a list of consensus reads of a particular gene family.
2. Sort the input list according to the HMM start coordinates in the increasing order.
3. Find a consensus read representing the starting point and add it into the *working set*.
4. The neighborhood of the *working set* consist of at most two consensus reads, one that is the closest on the left ($L$) and one that is the closest on the right ($R$) of the *working set*. We denote these consensus reads as $L$ and $R$, respectively (Fig. 7).
5. For each consensus sequence $S$ from the *working set* and for each pair ($L$, $S$) and ($S$, $R$), compute:
   a. overlap probability
   b. expected overlap length
   c. overlap score
6. If there is a sufficient overlap between at least one pair ($L$, $S$) or ($S$, $R$), the pair with the highest overlap score is chosen, as defined in (Section 2.4). Let ($L$, $S$) be the pair with the highest overlap. Remove $S$ from the *working set*. Join ($L$, $S$) into a consensus sequence (i.e. a super-read), as defined in (Section 2.3) and add it into the *working set*. Redefine $L$, as the first consensus read on the left of $L$. If ($S$, $R$) is the pair with the highest score, proceed analogously.
7. If there is no sufficient overlap found at step (6), add $L$ and $R$ into the *working set* and redefine $L$ and $R$ in the same way as in (6).
8. If the neighborhood is not empty, i.e. $L$ or $R$ was redefined at step (6) or (7), go to step (5). If $L$ or $R$ cannot be redefined, it is not considered in the next steps of the algorithm.
9. Output super-reads as annotated contigs.

In the algorithm, a consensus sequence is represented via a probability matrix as described in (Section 2.3). Mismatching bases within a sufficient overlap most likely represent a substitution error, where one of the bases has a relatively low quality score, thus, the base with a higher quality score corrects such a substitution error. Substitutions representing genuine strain variation are represented by overlap positions with different bases with relatively high quality scores, therefore, such overlaps of consensus reads representing different strains almost never pass the minimum required overlap probability threshold. Consensus reads containing insertion or deletion errors have very low overlap probabilities with other consensus reads or super-reads and are therefore unlikely to be joined into longer consensus sequences, thus super-read positions with coverage of at least two are mostly error-corrected in terms of insertion and deletion sequencing errors.

## 3. Results

We have evaluated *Snowball* using ten simulated datasets, each containing three closely related *E.Coli* strains (Section 3.4). Our aim was to answer the following questions: Were the contigs assembled correctly? How long are the resulting contigs? Did the assembly recover the reference strain sequences from which the input paired-end reads were generated? As a reference method, we have used the *SAT* assembler (Zhang *et al.*, 2014), since this is to our knowledge the only currently available reference-free gene assembler of gene domains of interest for metagenomic data.

### 3.1. Per-base error

We have computed the per-base error for all assembled contigs of all simulated datasets (Fig. 8). For each contig, we have determined the reference strain sequence and coordinates of a particular contig sequence within a respective reference sequence from which it originates. The per-base error is defined as the percentage of bases that differ between a contig sequence and the respective sub-sequence of the reference sequence, i.e. it corresponds to the hamming distance between the two sequences. Note, that closely related strains share large sequence regions; therefore a contig can be well mapped onto several reference sequences of distinct strains. In this case, a reference sequence, onto which a contig maps with the lowest hamming distance, is considered to be the reference strain sequence from which it originates. If a contig maps onto several sequences of different strains, with exactly the same error, we consider it to originate from all these strains. The coverage of a contig position is equal to the number of reads covering a respective position. In the "*Snowball*" algorithm, we keep track of all consensus reads that a contig consists of. For the *SAT* assembler, we have used *BWA* (Li and Durbin, 2009) to map consensus reads onto the contigs. We have computed the per-base error for each coverage [2,...,20] separately. Low-coverage positions have typically higher per-base error, as there is not enough information available to correct sequencing errors. This is most pronounced at positions with coverage one, where the per-base error corresponds to the substitution error of a respective sequencing platform (i.e. ~2.37% for our simulated datasets). At positions with higher coverage, the error-correction mechanism build-in the "*Snowball*" algorithm yields very low (~0.02%) per-base error (Fig. 8a). For the *SAT* assembler, contig positions with high-coverage correspond to consensus sequences containing reads of several strains, which yields relatively high per-base error (Fig 8b). This proves that the error-correction incorporated in the "*Snowball*" algorithm is indispensable for the assembly of closely related strains.

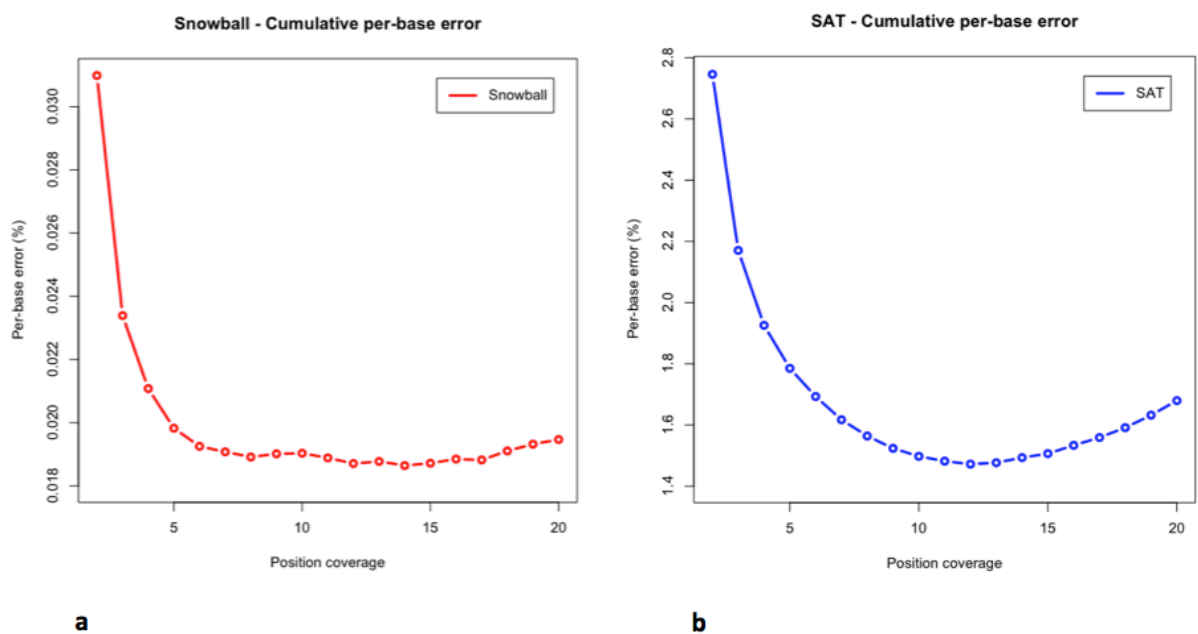

a   b

**Figure 8.** Cumulative per-base error.
Cumulative per-base error for the *Snowball* (panel a) and *SAT* (panel b) assemblers. We have computed the per-base error in a cumulative way, i.e. for $X \in [2, \ldots, 20]$ (on the horizontal $x$-axes), $Y$ (on the vertical $y$-axes) is equal to the per-base error at contig positions with coverage greater or equal to $X$. Note the different scale on the $y$-axes.

### 3.2. Relative contig length

We have computed the average number of assembled contigs and the average cumulative length of all contigs (in Mb) per simulated dataset (Fig. 9). As the assembled contigs should cover the full length of the respective gene sequences sufficiently well, we have aligned each contig to the respective profile HMM and computed the fraction of the model (i.e. the corresponding multiple sequence alignment) it covers. For each contig, this gave us, an estimate of its relative length with respect to the particular profile HMM. We have used this information to compute the results, e.g. using only longer contigs covering at least 50% (60%, 70%, etc.) of respective profile HMMs. The results show that *Snowball* produced substantially more and longer contigs than the *SAT* assembler.

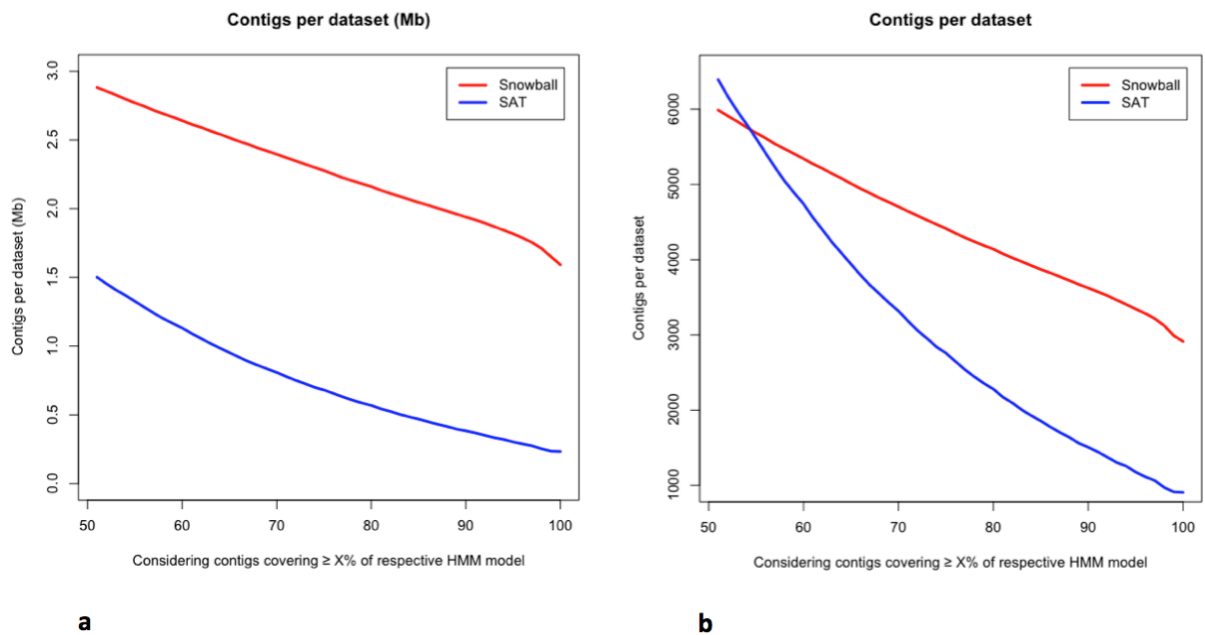

a                                                                b

**Figure 9.** Contigs per dataset.
Cumulative contig length per dataset, considering only contigs covering ≥ $X\%$ of respective profile HMM (panel *a*). Average number of contigs per dataset, considering only contigs covering ≥ $X\%$ of respective profile HMM (panel *b*). Here, the variable $X$ corresponds to the values on the (horizontal) $x$-axes of the graphs.

### 3.3. Reference coverage

We have computed, what part of the reference strain sequences, from which the input reads were generated, was recovered by the assembled contigs, per dataset on average (Fig. 10). As explained in (Section 3.1), assembled contigs can be mapped onto one or more reference strain sequences with the same minimum hamming distance. We consider that a contig covers all the reference strain sequences, onto which it can be mapped with exactly the same minimum per-base error. Positions of reference sequences that are covered by at least one contig are denoted as covered positions. For each dataset, we have computed the number and percentage of the covered positions. Moreover, as explained in (Section 3.2), we have computed these measures for contigs covering ≥ *X%* of respective profile HMMs (where the variable *X* corresponds to the values on the *x*-axes of the graphs). The overall relatively low coverage of the reference sequences can be explained by low sequencing coverage of some of the reference strain sequences (Section 3.4). Also, as we only assemble coding sequences of the reference strain sequences, for which we have used profile HMMs as the input, regions of the reference strain sequences that are not covered by the profile HMMs remain unassembled. Nevertheless, we have shown that *Snowball* recovered substantially more reference strain sequences than the *SAT* assembler.

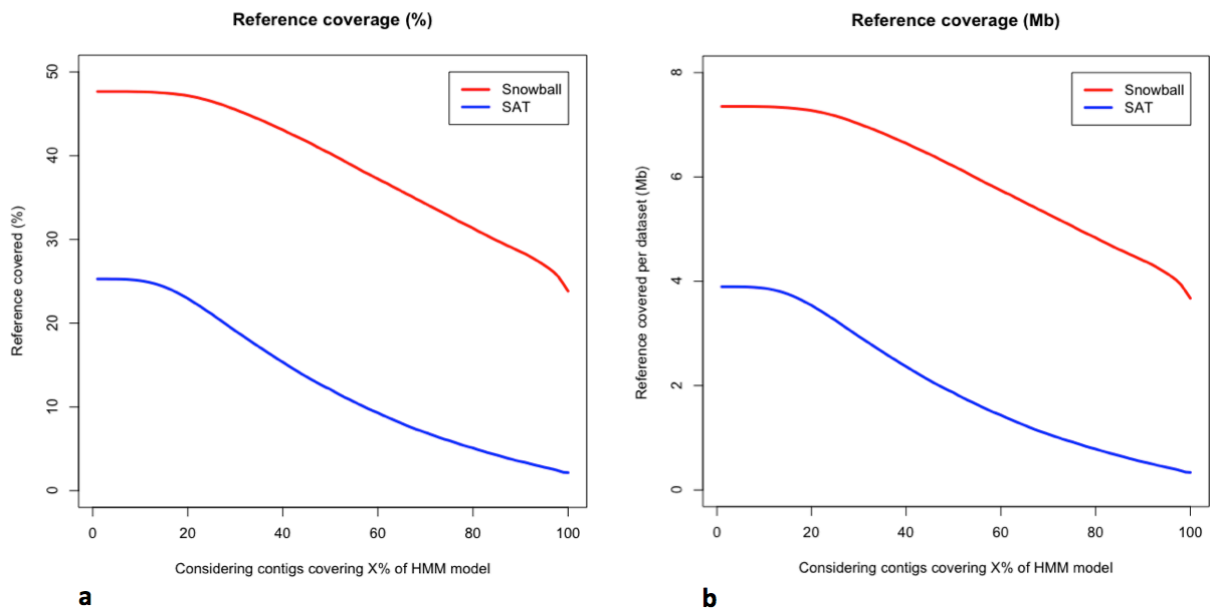

**Figure 10.** Coverage of the reference strain sequences.
Percentage of the recovered reference strains, per dataset on average, considering only contigs covering ≥ *X%* of respective profile HMMs (panel *a*). Corresponding absolute values (Mb) are depicted in (panel *b*). The variable *X* corresponds to the values on the *x*-axes.

## 3.4. Simulated datasets details

We have based our evaluation on ten simulated datasets, each containing three closely related *E.Coli* strains (Table 1). The strain abundances correspond to randomly drawn numbers from the log-normal distribution (mean=1, standard deviation=2), where the numbers were limited to interval [1, ..., 50], to avoid both data explosion and extremely low strain abundances. The *ART* (Huang *et al.*, 2012) read simulator (version 2.3.6) was employed to generate Illumina HiSeq 2500 paired-end reads (read length=150bp, mean insert size=225, standard deviation=23), where the strain coverage used for the read simulation corresponds to the strain abundance. We have used the same read simulator and strategy to generate strain abundances, which was used in the (CAMI challenge, 2015). Strain abundance of a particular strain thus tells us with which coverage the strain sequence of a simulated dataset was sequenced. We have used the default *ART* Illumina HiSeq 2500 empirical error profile, which yields reads with ~2.37% substitution error. Note, that we have chosen *E.Coli* strains for the evaluation, since there are more than 745 *E.Coli* (draft) genomes available in the public databases, from which we were able to generate many simulated datasets. For each dataset, we also provide per-dataset results (Table 1, Sections 3.1–3.3) that show that *Snowball* performed substantially better than the *SAT* assembler for all simulated datasets.

| Dataset | (a)Dataset Strains (Accessions) | (b)Strain coverage | (c)Per-base error at position coverage ≥ 5 | | (d)Contig length (Kb) 70% HMM model | | (e)Ref. cov. 70% HMM model | |
|---|---|---|---|---|---|---|---|---|
| | | | Snowball | SAT | Snowball | SAT | Snowball | SAT |
| 1 | NZ_AKLX00000000 NZ_ANLR00000000 NZ_AKLB00000000 | 25.6 06.2 02.6 | 0.022% | 1.627% | 2,845 | 1,113 | 41.1% | 9.7% |
| 2 | NZ_AEZU00000000 NZ_AIGR00000000 NZ_AIGN00000000 | 34.7 32.2 05.3 | 0.047% | 1.919% | 3,391 | 2,811 | 39.9% | 16.1% |
| 3 | NZ_AIGZ00000000 NZ_AKNI00000000 NZ_AIHC00000000 | 15.2 01.9 08.2 | 0.021% | 1.674% | 2,691 | 615 | 35.0% | 6.0% |
| 4 | NC_012759 NZ_AIFV00000000 NZ_AMTH00000000 | 13.3 07.7 13.5 | 0.023% | 1.749% | 2,841 | 1,003 | 40.1% | 10.2% |
| 5 | NZ_AIFA00000000 NZ_AIFD00000000 NZ_AIEZ00000000 | 03.0 01.9 02.2 | 0.010% | 1.845% | 1,122 | 40 | 15.7% | 0.5% |
| 6 | NZ_AIHP00000000 NZ_AIHO00000000 NZ_AIHS00000000 | 15.5 12.2 07.8 | 0.009% | 1.644% | 2,788 | 1,101 | 45.7% | 12.3% |
| 7 | NZ_AIHF00000000 NZ_AIHJ00000000 NZ_AIHH00000000 | 03.6 03.2 09.4 | 0.004% | 1.661% | 2,218 | 246 | 35.6% | 3.0% |
| 8 | NC_000913 NZ_AFAE00000000 NC_007779 | 04.3 10.6 01.5 | 0.017% | 1.708% | 2,194 | 240 | 31.6% | 2.4% |
| 9 | NZ_AIHN00000000 NZ_AIHM00000000 NZ_AIHL00000000 | 21.7 06.6 02.2 | 0.011% | 1.716% | 2,743 | 876 | 39.4% | 8.4% |
| 10 | NC_017660 NC_017635 NC_017664 | 02.1 02.1 02.8 | 0.005% | 1.521% | 1,118 | 28 | 17.0% | 0.3% |
| Average over all datasets | | | 0.020% | 1.785% | 2,395 | 808 | 34.3% | 7.0% |

**Table 1**. Simulated datasets overview.
(column a) Accession numbers of individual *E.Coli* strains of simulated datasets. (column b) Strain coverage of respective strains in the datasets. (column c) Per-base error at contig positions with coverage ≥ *5* (Fig. 8). (column d) Cumulative contig length at *X = 70* of (Fig. 9a). (column e) Percentage of recovered data at *X = 70* of (Fig. 10a).

## 4. Conclusions

We have presented *Snowball*, a novel strain aware gene assembler for gene domains of interest of the shotgun metagenomic data. *Snowball* performs gene assembly of individual gene domains based on read overlaps and error-correction using read quality scores at the same time, which result in very low per-base error rates. Our method uses profile HMMs to guide the assembly. Nonetheless, the method is reference-free, as it does not require closely related reference genomes of the studied species to be available. We have assessed the performance of *Snowball* using ten simulated datasets, each containing three closely related *E.Coli* strains. We have compared our *Snowball* assembler to the *SAT* assembler, which, to our knowledge, establishes the current state of the art in gene assembly. The results showed that *Snowball* had substantially lower per-base error, assembled more and longer contigs and recovered more data from the input paired-end reads. We have shown that the incorporation of the error-correction mechanism is indispensible for the assemblies of closely related strains. To our knowledge, *Snowball* is the first strain aware reference-free gene assembler, as the assembly of closely related strains is still often a challenging task for most of the current assemblers, including the *SAT* assembler. We believe that our tool will be valuable for researchers studying species evolution (e.g. genes under selection) and strain or gene diversity (e.g. virulent genes). *Snowball* is implemented in Python, runs on a user-defined number of processor cores in parallel, runs on a standard laptop and can be easily installed under Linux or OS X.